\documentclass[10pt]{article}
\usepackage{graphicx}
\usepackage{amssymb,amsfonts,amsmath}
\usepackage{color}
\usepackage{cite}
\usepackage{bm}

\usepackage{setspace} 
\usepackage[labelfont=bf,labelsep=period,justification=raggedright]{caption}

\makeatletter
\renewcommand{\@biblabel}[1]{\quad#1.}
\makeatother

\date{}

\newcommand{\EQ}{Eq.~}
\newcommand{\EQS}{Eqs.~}
\newcommand{\FIG}{Fig.~}

\newcommand{\MbG}{\overline{b}_{\rm G}}
\newcommand{\MbB}{\overline{b}_{\rm B}}

\begin{document}

\begin{flushleft}
{\Large \textbf{Coevolution of trustful buyers and cooperative sellers in the trust game}
}
\\
Naoki Masuda$^{1,2,\ast}$ and Mitsuhiro Nakamura$^{1}$
\\
\bf{1} Department of Mathematical Informatics,
The University of Tokyo,
Bunkyo, Tokyo, Japan
\\
\bf{2} PRESTO, Japan Science and Technology Agency,
Kawaguchi, Saitama, Japan
\\
$\ast$ E-mail: masuda@mist.i.u-tokyo.ac.jp
\end{flushleft}

\section*{Abstract}
Many online marketplaces enjoy great success. Buyers and sellers in successful markets carry out cooperative transactions even if they do not know each other in advance and a moral hazard exists. An indispensable component that enables cooperation in such social dilemma situations is the reputation system. Under the reputation system, a buyer can avoid transacting with a seller with a bad reputation. A transaction in online marketplaces is better modeled by the trust game than other social dilemma games, including the donation game and the prisoner's dilemma. In addition, most individuals participate mostly as buyers or sellers; each individual does not play the two roles with equal probability. Although the reputation mechanism is known to be able to remove the moral hazard in games with asymmetric roles, competition between different strategies and population dynamics of such a game are not sufficiently understood. On the other hand, existing models of reputation-based cooperation, also known as indirect reciprocity, are based on the symmetric donation game. We analyze the trust game with two fixed roles, where trustees (i.e., sellers) but not investors (i.e., buyers) possess reputation scores. We study the equilibria and the replicator dynamics of the game. We show that the reputation mechanism enables cooperation between unacquainted buyers and sellers under fairly generous conditions, even when such a cooperative equilibrium coexists with an asocial equilibrium in which buyers do not buy and sellers cheat. In addition, we show that not many buyers may care about the seller's reputation under cooperative equilibrium. Buyers' trusting behavior and sellers' reputation-driven cooperative behavior coevolve to alleviate the social dilemma.

\section*{Introduction}\label{sec:introduction}

The number of transactions executed in online marketplaces such as
eBay is soaring up.  To buy a desired item, a buyer must first trust
in a seller by paying in advance.  Because the seller may lose
little by dismissing a single buyer, the seller may be tempted to ship
a counterfeit item or may not even
transport the purchase to the buyer.
If many sellers
behave maliciously toward buyers, the marketplace would collapse.  Here
is a moral hazard. A classical example in which counterfeit items
would prevail is the ``market for lemons'' or the used car market
\cite{Akerlof1970QJE}.  Nevertheless, many auction sites and related online services,
including opinion forums, price comparison sites, and product review sites,
enjoy prosperity without
seriously being swamped by the malicious behavior of users
\cite{Resnick2000ACM,Malaga2001ECR,Dellarocas2003ManSci,Brown2006California}.

The main mechanism to elicit the cooperative behavior of sellers in such a social dilemma situation is the
online reputation management system, also called the feedback
mechanism
\cite{Resnick2000ACM,Malaga2001ECR,Dellarocas2003ManSci,Brown2006California}.
When a reputation management system is implemented, a buyer is invited
to evaluate the seller after a transaction so that other buyers can
refer to the reputation of this seller in the future.
A seller with a good overall
reputation would successfully sell items to many buyers in the long run,
whereas a seller with a bad reputation would be avoided by buyers.
In a seminal paper,
Klein and Leffler analyzed the role of reputations in alleviating
a moral hazard
\cite{Klein1981JPE}.

Reputation mechanisms in online marketplaces are often cited as
a successful example of indirect reciprocity
\cite{Bolton2004ManSci,Nowak2005nature}. Indirect
reciprocity (precisely, downstream
or vicarious reciprocity as its major subtype 
\cite{Nowak2005nature,Sigmund2010book})
is a mechanism for the alleviation of social dilemmas. It
dictates that an individual $i$ with a good reputation is helped by another
individual that $i$ has not met. $i$ may also help other individuals that
$i$ does not know. In fact, most buyer-seller pairs
conduct only one transaction on eBay, such that a buyer probably
does not know
the seller with whom the buyer is about to transact \cite{Resnick2002AAM}.
It is theoretically established that indirect reciprocity
enables cooperation in social dilemma games
under proper conditions
\cite{Nowak1998nature,Nowak1998jtb,Leimar2001RoyalB,Panchanathan2003jtb,OhtsukiIwasa2004jtb,Brandt2004jtb,Nowak2005nature,Sigmund2010book}.

However, the mechanism governing 
cooperation between unacquainted buyers and sellers
observed in real online transactions does not resemble that provided
by these models.
The existing models of indirect reciprocity are mostly based on the
donation game, which is a type of social dilemma games.
In the donation game, two players are chosen
from a population, one as donor and the other as recipient.
If the donor helps the recipient, the recipient
gains a benefit, which is larger than the cost that the donor is charged.
If the donor does not help the recipient, the donor and
the recipient gain nothing.  The donor's help contributes to social
welfare, whereas the donor is better off withholding the
help. 

In the previous models of indirect reciprocity
\cite{Nowak1998nature,Nowak1998jtb,Leimar2001RoyalB,Panchanathan2003jtb,Brandt2004jtb,Brandt2006jtb,OhtsukiIwasa2004jtb,OhtsukiIwasa2006jtb,OhtsukiIwasa2007jtb,Nowak2005nature,Chalub2006jtb,Pacheco2006PlosComputBiol,Uchida2010jtb,Sigmund2010book}, 
each player is selected as many times as donor and
recipient per generation (we discuss an exception
\cite{Johnstone2007royalb} in the Discussion).
Therefore, the players are essentially
involved in a symmetric prisoner's dilemma game. In contrast, the social
dilemma game effectively played in online marketplaces 
is a highly asymmetric
game. Buyers and sellers are distinct roles that each individual does
not play with equal probability \cite{Resnick2002AAM}.  In addition,
in the donation game and the symmetric prisoner's dilemma, there is no
concept wherein one player invests trust in a peer player in the
one-shot game. Theoretical models of reputation-based cooperation using different
social dilemma games such as the ultimatum game also assume
symmetrization of the two roles
\cite{Nowak2000Science,Sigmund2001PNAS}.  This is also
the case for the models of
reputation-based cooperation analyzed in economic literature
\cite{Raub1990AJS,Kandori1992RES}.

The trust game \cite{Dasgupta1988book,Kreps1990chap,Berg1995geb,Fehr2003Nature} seems to be a much better model for
online marketplaces \cite{Bolton2004ManSci}.
As shown in \FIG\ref{fig:schem trust
  game}, the buyer, also called the investor, first decides whether to buy
an item from the seller. If the buyer buys,
the seller, also called the trustee, decides whether to
ship the appropriate item to the buyer.  
If the buyer buys and the seller does not ship, succumbing to the
temptation to defect, the seller gains the largest payoff 1, and the
buyer gains the smallest payoff $-1$.
If the buyer buys and the seller
ships, both players obtain a relatively large payoff $r$, where $0<r<1$.
  If the buyer does not buy, both
players gain a relatively small payoff 0, which is the unique
Nash equilibrium of the game.
A cooperative society is realized when
the buyer buys and the seller ships.
In laboratory experiments, 
humans cooperate to some extent in the trust game
\cite{Berg1995geb,McCabe2000pnas,McCabe2003JEBO,Fehr2003Nature}, and reputation mechanisms
enhance cooperation
\cite{Keser2003IBM,Bolton2004ManSci,Basu2009pnas,Bracht2009jpe}.
The trust game with a reputation mechanism also approximates the situation of
commerce conducted by the medieval
Maghribi traders \cite{Greif1993AER,Greif2006book} and the market for lemons (if
a gossip-based reputation mechanism is operational)
\cite{Akerlof1970QJE}.

It was shown in a seminal paper that
cooperation based on reputations is
possible in the essentially asymmetric trust game
\cite{Klein1981JPE}.
Nevertheless, several important questions in this framework
are theoretically unresolved. 
Is cooperation more ubiquitous than uncooperation, in particular
when these two situations are both stable
equilibria? How do different reputation-based strategies of buyers
compete in a population? 
How does the reputation-based cooperation emerge
through population dynamics?
Non-evolutionary
numerical results for the trust game with
reputation mechanisms
\cite{Diekmann2005ESSA,Resnick2006ExpEcon} do
not explain the stability and emergence of cooperation.  An evolutionary
theory \cite{Sigmund2001PNAS,Sigmund2010book} and numerical
simulations \cite{Bravo2008Rational} showed that reputations induce
cooperation in the trust game.  However, in these papers, each player
serves the two roles with equal probability such that the game is essentially
the symmetric prisoner's dilemma.

We theoretically clarify the possibility of
reputation-based cooperation in the trust game by analyzing the
Nash equilibria and the coevolutionary
replicator dynamics of buyers and sellers. 
We show that coevolution of cooperative buyers and
sellers is realized relatively easily.
In particular, the fraction of buyers that score sellers does not have to be
large, and many buyers do not discriminate between good and bad sellers even
when cooperation prevails.

\section*{Model}

\subsection*{Trust game}\label{sub:trust game}

We analyze equilibria and evolutionary dynamics of the trust game
(\FIG\ref{fig:schem trust game}) in an infinite population of four
types of buyers and two types of sellers under a reputation
mechanism.
Each seller possesses the binary reputation, good (G) or bad (B), that
dynamically changes as a result of the single-shot trust game.
Referring to the two reputations as G and B is purely conventional.

In a time unit, each seller $i_{\rm s}$ plays the trust game
with a randomly chosen buyer $i_{\rm b}$. 
The buyer $i_{\rm b}$ first decides whether to buy
an item from $i_{\rm s}$ possibly on the basis of $i_{\rm s}$'s reputation.
If $i_{\rm b}$ does not buy, no transaction occurs, leaving the
payoff 0 to both $i_{\rm b}$ and $i_{\rm s}$.  If $i_{\rm b}$
buys, $i_{\rm s}$ decides
whether to ship the item (cooperate; C) or not (defect; D). If $i_{\rm s}$
ships the item, then both $i_{\rm b}$ and $i_{\rm s}$ obtain $r$. If $i_{\rm s}$
does not ship the item and exploits $i_{\rm b}$, $i_{\rm s}$ gains 1, and
$i_{\rm b}$ gains $-1$.  If $r>1$, it is beneficial for both players to
cooperate. If $r<0$, transaction would never occur irrespective of the
reputation mechanism. We set $0<r<1$ to represent the social dilemma.

We repeat the procedure explained above for $T$ time units such that
each player plays the trust game $T$ times on an average.
We assume that $1\ll T\ll N$ such that the probability that the same
pair of buyer and seller interact more than once within time
$T$ is infinitesimally small.

\subsection*{Social norms}

When $i_{\rm b}$ decides to buy, $i_{\rm b}$ assigns
G or B to $i_{\rm s}$ on the basis of $i_{\rm s}$'s action.
The new reputation of $i_{\rm s}$ is
instantaneously propagated to the entire population such that any
buyer can refer to this information when playing with $i_{\rm s}$
in later times.
We refer to the rule according to
which the buyer assigns a reputation to the seller 
as a social norm.

An intuitively rational social norm, which reputation mechanisms
in successful online marketplaces apply,
is to assign G and B when $i_{\rm s}$ has
cooperated and defected, respectively. This norm is called
image scoring \cite{Nowak1998nature,Nowak1998jtb}.  Alternatively,
$i_{\rm b}$ may assign G no matter whether $i_{\rm s}$ cooperates or defects.
We call this social norm indifferent scoring. 
A buyer is assumed to commit the assignment error
with probability $0<\mu< 1/2$ such that the seller receives a
reputation that is contrary to what is expected from the social norm. 

In fact, some but not all of the buyers may be interested
in rating sellers \cite{Resnick2000ACM,Malaga2001ECR,Dellarocas2003ManSci}.
To investigate this scenario, we consider
a case in which a buyer has a unique scoring type as well as strategy.
We assume that a fraction of
$\theta$ and $1-\theta$ buyers are
image scorers and indifferent scorers, respectively. 
We also assume that
 the scoring type does not affect the buyer's payoff, such that the fraction
of image scorers in the population does not change in the course of the replicator dynamics.
Alternatively, we can assume that
each buyer is a permanent indifferent scorer or
permanent image scorer. This assumption may be controversial.
We will return to this issue in 
Discussion.

We do not assume the reputation for buyers because in online
marketplaces, the impact of the seller's reputation is much larger than
that of the buyer's reputation \cite{Malaga2001ECR,Brown2006California}.
%
% buyer also submits feedback (Brown 2006 p.6)
%
Previous laboratory experiments that modeled the situation in auction
sites also neglected the reputation of buyers
\cite{Resnick2002AAM,Keser2003IBM,Bolton2004ManSci}.

\subsection*{Strategies}

The probability that the buyer decides to buy from G and B sellers
is denoted by $b_{\rm G}$ and $b_{\rm B}$, respectively.  We consider four
strategies for buyers: unconditional buyer (Buy) specified by $b_{\rm
  G}=b_{\rm B}=1-\epsilon$, where $0<\epsilon< 1/2$; discriminator
(Disc) specified by $b_{\rm G}=1-\epsilon$, $b_{\rm B}=\epsilon$;
anti-discriminator (AntiDisc) specified by $b_{\rm G}=\epsilon$,
$b_{\rm B}=1-\epsilon$; and unconditional no-buyer (NoBuy) specified
by $b_{\rm G}=b_{\rm B}=\epsilon$. 

As introduced in section ``Trust game'',
sellers have two strategies, C and D.
For simplicity, we
assume that the seller's action is deterministic, whereas
the buyer's action is stochastic. 

\section*{Results}

\subsection*{Payoffs}

The reputation of a seller obeys a Markov chain with two states
G and B. To obtain the payoffs,
we adopt the deterministic calculations developed by Ohtsuki and Iwasa
\cite{OhtsukiIwasa2004jtb,OhtsukiIwasa2006jtb,OhtsukiIwasa2007jtb}.

If $t$ ($0\le t\le T$) is sufficiently large, we can approximate
the probability of the G reputation for a C seller and that for a
D seller by the population averages
denoted by $\rho_{\rm C}$ and $\rho_{\rm D}$, respectively.
The dynamics of the reputation averaged over the C sellers and
that averaged over the D sellers are represented as
\begin{align}
\frac{d\rho_{\rm C}}{dt} = & -\rho_{\rm C}\left[\theta\MbG^{\rm IM} + (1-\theta)\MbG^{\rm IN} \right]\mu 
+ (1-\rho_{\rm C})\left[\theta\MbB^{\rm IM} + (1-\theta)\MbB^{\rm IN} \right](1-\mu),
\label{eq:drhoC/dt for general theta}\\
\frac{d\rho_{\rm D}}{dt} = & - \rho_{\rm D}\left[\theta\MbG^{\rm IM} (1-\mu) + (1-\theta)\MbG^{\rm IN}\mu
\right] + (1-\rho_{\rm D})\left[\theta\MbB^{\rm IM} \mu + (1-\theta)\MbB^{\rm IN}(1-\mu) \right],
\label{eq:drhoD/dt for general theta}
\end{align}
where
$\MbG^a$ and $\MbB^a$ ($a={\rm IN}$ for indifferent
scorer or ${\rm IM}$ for image scorer)
are the probabilities that the buyer of scoring type $a$
decides to buy from a G and B seller, respectively.
For example,
$\left[\theta\MbG^{\rm IM} + (1-\theta)\MbG^{\rm IN} \right]$ on the
right-hand side of \EQ\eqref{eq:drhoC/dt for general theta}
represents the probability that a buyer decides to buy from a C seller.
Because the multiplicative factor $\mu$ represents the probability
that the C seller that has cooperated mistakenly receives B reputation,
the first term on the right-hand side of \EQ\eqref{eq:drhoC/dt for
  general theta} represents the case where the reputation of C seller
turns from G to B. Similarly, the second term represents the case
where the reputation of C seller turns from B to G.
It should be noted that
the seller's reputation does not change when the buyer decides
not to buy.

The probabilities that the buyer decides to buy from a G and B seller
are respectively represented as
\begin{align}
\MbG^a=& (1-\epsilon)(y_1^a+y_2^a)+\epsilon (y_3^a+y_4^a),\label{eq:def of MbG a}\\
\MbB^a=& (1-\epsilon)(y_1^a+y_3^a)+\epsilon (y_2^a+y_4^a),\label{eq:def of MbB a}
\end{align} 
where $y_i^{\rm IN}$ and $y_i^{\rm IM}$
are 
the fractions of buyers with strategy $i$ among the indifferent scorers and
among the image scorers, respectively.
Buy, Disc, AntiDisc, and NoBuy correspond to
$i=1$, 2, 3, and 4, respectively;
 $\sum_{i=1}^4 y_i^{\rm IN} = \sum_{i=1}^4 y_i^{\rm IM} = 1$.

For simplicity, we assume that
\begin{equation}
y_i\equiv y_i^{\rm IN}=y_i^{\rm IM}\quad (1\le i\le 4)
\label{eq:yiIN=yiIM}
\end{equation}
is initially satisfied.
In other words, the scoring type and strategy are independent of each other.
Note that $0\le y_i\le 1$ ($1\le i\le 4$) and $\sum_{i=1}^4 y_i = 1$.
Because \EQS\eqref{eq:def of MbG a}, \eqref{eq:def of MbB a}, and \eqref{eq:yiIN=yiIM}
imply $\MbG\equiv\MbG^{\rm IN}=\MbG^{\rm IM}$ and $\MbB\equiv\MbB^{\rm IN}=\MbB^{\rm IM}$,
\EQS\eqref{eq:drhoC/dt for general theta} and \eqref{eq:drhoD/dt for
  general theta} give the limit values
\begin{align}
\rho_{\rm C}^* = & \frac{(1-\mu)\MbB}{\mu\MbG + (1-\mu)\MbB},
\label{eq:reputation C seller}\\
\rho_{\rm D}^* = & \frac{c_2\MbB}{c_1\MbG+c_2\MbB},
\label{eq:reputation D seller}
\end{align}
where
\begin{align}
\MbG=& (1-\epsilon)(y_1+y_2)+\epsilon (y_3+y_4),\label{eq:def of MbG}\\
\MbB=& (1-\epsilon)(y_1+y_3)+\epsilon (y_2+y_4),\label{eq:def of MbB}
\end{align} 
and we set 
\begin{align}
c_1 =& \theta(1-\mu)+(1-\theta)\mu,\label{eq:c1}\\
c_2 =& \theta\mu+(1-\theta)(1-\mu),\label{eq:c2}
\end{align}
for notational convenience.
Equation~\eqref{eq:reputation C seller} does not depend on $\theta$,
which reflects the fact that
both the indifferent and image scorers evaluate C sellers as
G with a large probability $1-\mu$ ($>1/2$). 
In contrast, \EQ\eqref{eq:reputation D seller}
implies that $\rho_{\rm D}^*$ decreases with $\theta$. This is because
the image scorer may issue a B reputation with a large probability, 
whereas the indifferent scorer does not.

For sufficiently large $T$,
the reputations are in the equilibrium
almost all the time. 
Then, the buyer's payoff should be equal to
\begin{align}
P_i^{\rm b}(b_{\rm G}, b_{\rm B}) = &
rx \left[\rho_{\rm C}^* b_{\rm G} + (1-\rho_{\rm C}^*) b_{\rm B}\right]
- (1-x) \left[\rho_{\rm D}^* b_{\rm G} + (1-\rho_{\rm D}^*) b_{\rm
    B}\right]\notag \\
 = & \frac{rx\left[(1-\mu)b_{\rm G}\MbB +
    \mu\MbG b_{\rm B}\right]} {\mu\MbG + (1-\mu)\MbB}
 - \frac{(1-x)(c_2 b_{\rm G}\MbB + c_1\MbG b_{\rm B})}
{c_1\MbG+c_2\MbB},
\label{eq:payoff buyer}
\end{align}
where
\begin{equation}
(b_{\rm G},b_{\rm B})=\begin{cases}
(1-\epsilon,1-\epsilon), & \text{if } i=1 \text{ (Buy)},\\
(1-\epsilon,\epsilon), & \text{if } i=2 \text{ (Disc)},\\
(\epsilon,1-\epsilon), & \text{if }  i=3 \text{ (AntiDisc)},\\
(\epsilon,\epsilon), & \text{if } i=4 \text{ (NoBuy)},
\end{cases}
\end{equation}
and $x$ ($0\le x\le 1$) is the fraction of C sellers.
The payoffs for
C and D sellers are given by
\begin{equation}
P^{\rm s}_{\rm C} = r\left[\rho_{\rm C}^*\MbG
 + \left(1-\rho_{\rm C}^*\right)\MbB\right]
= \frac{r\MbG\MbB}{\mu\MbG + (1-\mu)\MbB}
\label{eq:payoff C seller}
\end{equation}
and
\begin{equation}
P^{\rm s}_{\rm D} =
\rho_{\rm D}^*\MbG + \left(1-\rho_{\rm D}^*\right)\MbB
= \frac{\MbG\MbB}{c_1\MbG+c_2\MbB},
\label{eq:payoff D seller}
\end{equation}
respectively. 

Even if we consider the stochastic dynamics of the reputation and
the buyer's action, the values of the payoffs derived above
give the precise mean values.

\bigskip

\textbf{Proposition 1:} Regardless of the initial reputation value of a seller,
in the limit $T\to\infty$ and $T/N\to 0$,
the expected payoff for the buyer is given by
\EQ\eqref{eq:payoff buyer} and that for the seller is given by
\EQS\eqref{eq:payoff C seller} and \eqref{eq:payoff D seller}.

\bigskip

\subsection*{Proof of Proposition 1}

The reputation score of the seller obeys a Markov chain
with two states. Consider a C seller $i_{\rm s}$ with reputation G.
The G reputation does not change in one time step if
the paired buyer $i_{\rm b}$ decides to buy with
probability $\MbG$ and correctly assign G to $i_{\rm s}$
with probability $1-\mu$ or, $i_{\rm b}$ decides not to buy
with probability $1-\MbG$. Otherwise, $i_{\rm s}$'s reputation
turns into B. When $i_{\rm s}$ has reputation B,
it is unchanged if $i_{\rm b}$ decides to buy
with probability $\MbB$ and commit assignment error
with probability $\mu$, or $i_{\rm b}$ decides not to buy
with probability $1-\MbB$. Otherwise,
$i_{\rm s}$'s reputation turns into G.

Therefore, the transition matrix of the Markov chain is represented as
\begin{equation}
M\equiv \begin{pmatrix}
(1-\mu)\MbG+(1-\MbG) & \mu\MbG\\
(1-\mu)\MbB & \mu\MbB+(1-\MbB)
\end{pmatrix},
\end{equation}
where $M_{ij}$ ($1\le i, j\le 2$) represents the transition
probability
from state $i$ to state $j$, and we associate G and B with states 1
and 2, respectively. Because $M$ is a nondegenerate
(right) stochastic matrix, we can
decompose $M$ using the left and right eigenvectors corresponding to
eigenvalue 1 as
\begin{equation}
M=\begin{pmatrix}1\\ 1\end{pmatrix}
\begin{pmatrix} \frac{(1-\mu)\MbB}{\mu\MbG+(1-\mu)\MbB} 
& \frac{\mu\MbG}{\mu\MbG+(1-\mu)\MbB} \end{pmatrix}
+ \lambda \bm u \bm v,
\end{equation}
where $-1<\lambda<1$ is the other eigenvalue of $M$, and 
$\bm v$ and $\bm u$ are the left and right eigenvectors corresponding
to $\lambda$, respectively. We do not calculate $\lambda$, $\bm v$,
and $\bm u$ because their values are immaterial in the following arguments.
Note that the eigenvectors are normalized such that
\begin{align}
\begin{pmatrix}\frac{(1-\mu)\MbB}{\mu\MbG+(1-\mu)\MbB} &
\frac{\mu\MbG}{\mu\MbG+(1-\mu)\MbB}\end{pmatrix}
\begin{pmatrix}1\\ 1\end{pmatrix} =
\bm v\bm u =& 1,\\
\bm v \begin{pmatrix}1\\ 1\end{pmatrix}=
\begin{pmatrix}\frac{(1-\mu)\MbB}{\mu\MbG+(1-\mu)\MbB} &
\frac{\mu\MbG}{\mu\MbG+(1-\mu)\MbB}\end{pmatrix}\bm u =& 0.
\end{align}

Assume that the C seller $i_{\rm s}$ initially 
has reputation G and B with probability $p_{\rm G, init}$ and
$p_{\rm B, init}$, respectively, where $p_{\rm G, init} + p_{\rm B,
  init}=1$. Then, the probability that $i_{\rm s}$'s reputation is G
and B after playing $t$ games
is given by the first and second columns
of $(p_{\rm G, init}\; p_{\rm B, init})M^t$, respectively.

The expected payoff for $i_{\rm s}$ in a single game
is equal to $r$ multiplied by
the probability that $i_{\rm b}$ decides
to buy. Therefore, the expected payoff for $i_{\rm
s}$ per single game, averaged over $1\le t\le T$,
converges in the limit
$T\to\infty$ to
\begin{align}
& \lim_{T\to\infty}\frac{1}{T}
\begin{pmatrix} p_{\rm G, init} & p_{\rm B, init}\end{pmatrix}
\left(I+M+M^2+\cdots+M^{T-1} \right)
\begin{pmatrix}r\MbG\\ r\MbB\end{pmatrix}\\
=&
\lim_{T\to\infty}\frac{1}{T}
\begin{pmatrix} p_{\rm G, init} & p_{\rm B, init}\end{pmatrix}
\left[T\begin{pmatrix}1\\1\end{pmatrix}
\begin{pmatrix} \frac{(1-\mu)\MbB}{\mu\MbG+(1-\mu)\MbB} & 
\frac{\mu\MbG}{\mu\MbG+(1-\mu)\MbB} \end{pmatrix}
+ \frac{1-\lambda^T}{1-\lambda} \bm u \bm v \right]
\begin{pmatrix}r\MbG\\ r\MbB\end{pmatrix}\\
=&
\begin{pmatrix} p_{\rm G, init} & p_{\rm B, init}\end{pmatrix}
\begin{pmatrix}1\\1\end{pmatrix}
\begin{pmatrix} \frac{(1-\mu)\MbB}{\mu\MbG+(1-\mu)\MbB} & 
\frac{\mu\MbG}{\mu\MbG+(1-\mu)\MbB} \end{pmatrix}
\begin{pmatrix}r\MbG\\ r\MbB\end{pmatrix}
= \frac{r\MbG\MbB}{\mu\MbG+(1-\mu)\MbB},
\end{align}
which reproduces \EQ\eqref{eq:payoff C seller}.
The expected payoff for the D seller, given by \EQ\eqref{eq:payoff D seller},
can be derived in the same manner.

Next, we calculate the payoff for a Buy buyer $i_{\rm b}$.
In each time step, the expected number of
C seller with reputation G and B with whom
$i_{\rm b}$ is paired is equal to
the first and second columns of
$x (p_{\rm G, init}\; p_{\rm B, init})M^t$, respectively.
When the C seller
$i_{\rm s}$ has reputation G (B), the expected payoff for $i_{\rm b}$
in a single game is equal to $r\MbG$ ($r\MbB$).
Therefore, the contribution of the C seller to
the expected payoff for $i_{\rm b}$ per single game,
averaged over $T$ games, converges to
\begin{equation}
\lim_{T\to\infty}\frac{1}{T} x
\begin{pmatrix} p_{\rm G, init} & p_{\rm B, init}\end{pmatrix}
\left(I+M+M^2+\cdots+M^{T-1} \right)
\begin{pmatrix}r\MbG\\ r\MbB\end{pmatrix}
= \frac{xr\MbG\MbB}{\mu\MbG+(1-\mu)\MbB}.
\label{eq:payoff Buy from C seller}
\end{equation}
This quantity
is equal to the first term on the right-hand side of
\EQ\eqref{eq:payoff buyer} when $b_{\rm G}=b_{\rm B}=1-\epsilon$
(i.e., Buy).
Analogous calculations for the case of D seller yields the second term
on the right-hand side of \EQ\eqref{eq:payoff buyer}
when $b_{\rm G}=b_{\rm B}=1-\epsilon$.
The payoff for Disc, AntiDisc, and NoBuy can be calculated in a similar manner.
Here is the end of the proof.

\subsection*{Nash equilibria}

Based on the expected payoff determined by Proposition 1,
we identify the equilibria of the game.
In the analysis, we exploit the fact that
$P_i^{\rm b}(b_{\rm G}, b_{\rm B})$ is linear in $b_{\rm G}$ and $b_{\rm B}$.
There are three types of Nash equilibria. 
The so-called
uncooperative equilibrium is composed of NoBuy and the D seller.
In the so-called cooperative equilibrium, Buy and Disc are
mixed in the buyer's strategy and the probability of C is large
in the seller's strategy. 
The cooperative equilibrium corresponds to the situation in which
buyers and sellers do not repeatedly interact but trust each other on
the basis of the reputation mechanism. When it exists, it coexists
with the uncooperative equilibrium.
The other equilibrium appears only for a singular parameter set. Therefore,
we are not concerned with it in the later analysis.

\bigskip

\textbf{Proposition 2:}
The asymmetric trust game with a reputation mechanism whose expected
payoffs are defined by \EQS\eqref{eq:payoff buyer}, \eqref{eq:payoff
C seller}, and \eqref{eq:payoff D seller}
possesses the following
three types of Nash equilibria.

\begin{enumerate}
\item
Uncooperative equilibrium:
combination of NoBuy and $x^*=0$. This pure-strategy
Nash equilibrium is also strict.

\item
Cooperative equilibrium:
the mixture of Buy and Disc, with 
the probability of Buy and Disc being
\begin{equation}
y_1^*= \frac{(-\mu+rc_1)(1-\epsilon)+(1-\mu-rc_2)\epsilon}
{(1-\mu-rc_2)(1-2\epsilon)}
\label{eq:y1* cooperative equilibrium}
\end{equation}
and $y_2^*=1-y_1^*$, respectively. The probability of C seller
is given by 
\begin{equation}
x^* = \frac{c_1}{c_1+\mu}.
\label{eq:x* when Buy and Disc coexist repl}
\end{equation}
The cooperative equilibrium exists when
\begin{equation}
\frac{\mu(1-\epsilon)+(1-\mu)\epsilon}
{c_1(1-\epsilon) + c_2\epsilon}
<r<1.
\label{eq:Disc necessary cnd 1 replicated}
\end{equation}
The cooperative equilibrium is also asymptotically stable under the
replicator dynamics. For completeness,
the replicator dynamics of buyers and sellers are
given by
\begin{align}
\frac{{\rm d}y_i}{{\rm d}t}=& y_i
\left[P_i^{\rm b}(b_{\rm G}, b_{\rm B})-\overline{P^{\rm b}} \right],
\label{eq:dy/dt}\\
\frac{{\rm d}x}{{\rm d}t} =& x\left(P^{\rm s}_{\rm C}-P^{\rm s}_{\rm D}\right)\notag\\
=& x(1-x)\MbG\MbB
\left[\frac{r}{\mu\MbG+(1-\mu)\MbB} - \frac{1}{c_1\MbG+c_2\MbB} \right],
\label{eq:dx/dt}
\end{align}
respectively, where the buyer's mean payoff is given by
\begin{align}
\overline{P^{\rm b}} =& \left[\frac{rx}{\mu\MbG+(1-\mu)\MbB} - 
\frac{1-x}{c_1\MbG+c_2\MbB}\right]\MbG\MbB.
\end{align}

\item
A singular equilibrium:
combination of Disc buyer 
and a mixed seller's strategy with any $x$ satisfying
\begin{equation}
\frac{c_2}{1-\mu+c_2}\le x\le \frac{c_1}{\mu+c_1}.
\end{equation}
This equilibrium exists when
\begin{equation}
\frac{r}{\mu(1-\epsilon)+(1-\mu)\epsilon}
-\frac{1}{c_1(1-\epsilon)+c_2\epsilon}=0.
\label{eq:Disc singular main text}
\end{equation}

\end{enumerate}

We remark that the cooperative equilibrium is called so because
$\lim_{\mu\to 0} x^* = 1$. We prove Proposition 2 in the next section.

It should be noted that extending the concept of the evolutionary
stability to the asymmetric game is not straightforward.  In the
matrix game, a strictly (i.e., completely) mixed Nash equilibrium cannot
be an
asymptotically stable equilibrium under the replicator dynamics, and an
evolutionarily stable strategy in the asymmetric game is necessarily a
(pure) strict Nash equilibrium
\cite{Samuelson1992JET,Weibull1995book,Hofbauer1998book,Gintis2009book_evolving}.
Nevertheless, Proposition 2 dictates that the cooperative equilibrium
is an asymptotically stable strictly mixed strategy.  This is possible
because the payoff values are density-dependent in our model; it is
not a matrix game.  Therefore, we directly prove that the
cooperative equilibrium is asymptotically stable in the replicator
dynamics.

\subsection*{Proof of Proposition 2}

We identify all the mixed-strategy Nash
equilibria of the asymmetric game whose payoffs are given by
\EQS\eqref{eq:payoff buyer},
\eqref{eq:payoff C seller}, and \eqref{eq:payoff D seller}.

\subsubsection*{One buyer's strategy}

Consider a possible equilibrium composed of a single buyer's strategy.
If there is only Buy, AntiDisc, or NoBuy, we substitute $(\MbG,\MbB)
= (1-\epsilon,1-\epsilon)$, $(\epsilon,1-\epsilon)$, and
$(\epsilon,\epsilon)$, respectively,
in \EQS\eqref{eq:payoff C seller} and \eqref{eq:payoff D seller}
to obtain $P^{\rm s}_{\rm C} - P^{\rm s}_{\rm D}<0$ 
for $r<1$. Therefore, $x^*=0$ must be satisfied in
a possible Nash equilibrium.

When $x^*=0$, \EQ\eqref{eq:payoff buyer} is simplified to
\begin{equation}
P_i^{\rm b}(b_{\rm G}, b_{\rm B}) =
 - \frac{c_2 b_{\rm G}\MbB + c_1\MbG b_{\rm B}}
{c_1\MbG+c_2\MbB},
\label{eq:payoff buyer x^*=1}
\end{equation}
where $c_1$ and $c_2$ are defined in \EQS\eqref{eq:c1} and
\eqref{eq:c2},
respectively.
Equation~\eqref{eq:payoff buyer x^*=1} implies that
the payoff for NoBuy is larger than those for Buy, Disc, and AntiDisc.
Therefore, the
combination of NoBuy and D seller is the only (strict) Nash
equilibrium allowed in this regime. 
We call this equilibrium the uncooperative equilibrium.

Suppose instead that there is only Disc. If
\begin{equation}
\frac{r}{\mu\MbG+(1-\mu)\MbB}-\frac{1}{c_1\MbG+c_2\MbB}=0,
\label{eq:Disc singular}
\end{equation}
where $\MbG=1-\epsilon$ and $\MbB=\epsilon$,
we obtain $P^{\rm s}_{\rm C} = P^{\rm s}_{\rm D}$ for any $x$.
Substituting \EQ\eqref{eq:Disc singular} in \EQ\eqref{eq:payoff buyer}
yields
\begin{equation}
P_i^{\rm b}(b_{\rm G}, b_{\rm B}) =
\frac{\left[x(1-\mu)-(1-x)c_2\right]b_{\rm G}\MbB 
+ \left[x\mu-(1-x)c_1 \right]\MbG b_{\rm B}}
{c_1\MbG+c_2\MbB}.
\label{eq:payoff buyer Disc singular}
\end{equation}
Equation~\eqref{eq:payoff buyer Disc singular}
indicates that Disc obtains a payoff larger than or equal to
those of Buy, AntiDisc, and NoBuy
if
\begin{equation}
\frac{c_2}{1-\mu}\le \frac{x}{1-x}\le \frac{c_1}{\mu}.
\label{eq:Disc only range x}
\end{equation}
The range of $x$ that satisfies \EQ\eqref{eq:Disc only range x}
always exists because $\mu<1/2$ guarantees
$c_2/(1-\mu)<c_1/\mu$. Therefore, the combination of Disc and
\EQ\eqref{eq:Disc only range x}, i.e., 
\begin{equation}
\frac{c_2}{1-\mu+c_2}\le x\le \frac{c_1}{\mu+c_1},
\end{equation}
yields Nash equilibria.

If \EQ\eqref{eq:Disc singular} is not satisfied, which is
a generic case, only $x=0$ and $x=1$ may result in a Nash equilibrium
because the difference between
\EQS\eqref{eq:payoff C seller} and \eqref{eq:payoff D seller}
is independent of $x$. If $x^*=0$ (i.e., $P^{\rm s}_{\rm C}<P^{\rm
  s}_{\rm D}$), 
\EQ\eqref{eq:payoff buyer}
is simplified to
\begin{equation}
P_i^{\rm b}(b_{\rm G}, b_{\rm B}) = 
- \frac{c_2\epsilon b_{\rm G} + c_1(1-\epsilon) b_{\rm B}}
{c_1(1-\epsilon) + c_2\epsilon}.
\label{eq:B general theta 4}
\end{equation}
Equation~\eqref{eq:B general theta 4} implies that
NoBuy gains a larger payoff than Disc.
If $x^*=1$,
substituting
$\MbG=1-\epsilon$ and $\MbB=\epsilon$ in
\EQS\eqref{eq:payoff C seller} and \eqref{eq:payoff D seller}
and setting $P^{\rm s}_{\rm C} > P^{\rm s}_{\rm D}$
lead to the following necessary condition for Disc to be Nash:
\begin{equation}
r>\frac{\mu(1-\epsilon)+(1-\mu)\epsilon}
{c_1(1-\epsilon) + c_2\epsilon}.
\label{eq:Disc necessary cnd 1}
\end{equation}
Equation~\eqref{eq:Disc necessary cnd 1}
replicates \EQ\eqref{eq:Disc necessary cnd 1 replicated}.
When \EQ\eqref{eq:Disc necessary cnd 1} is
satisfied, substituting $x^*=1$, $\MbG=1-\epsilon$, and $\MbB=\epsilon$
in \EQ\eqref{eq:payoff buyer} yields
\begin{equation}
P_i^{\rm b}(b_{\rm G}, b_{\rm B}) =  \frac{r\left[(1-\mu)\epsilon b_{\rm G} +
    \mu (1-\epsilon) b_{\rm B}\right]} {\mu(1-\epsilon) +
  (1-\mu)\epsilon}.
\end{equation}
Therefore, Buy gains a larger payoff than Disc,
such that the pure Disc is not Nash.

In conclusion, the only pure (strict) Nash 
equilibrium is the uncooperative
equilibrium composed of NoBuy and D seller.

\subsubsection*{Mixture of two buyer's strategies}

Consider the mixed strategies (i.e.,
coexistence) of two buyer's strategies as candidates of Nash equilibria.
If we select two strategies out of Buy, AntiDisc, and NoBuy,
we can show $P^{\rm s}_{\rm C} < P^{\rm s}_{\rm D}$ in a manner 
similar to the case of the one buyer's strategy. In this case,
$x^*=0$ must hold true in the equilibrium.
Equation~\eqref{eq:payoff buyer} with $x^*=0$ indicates that 
the payoff for NoBuy is larger than that for
AntiDisc, which is larger than
that for Buy. Therefore,
such a mixed-strategy Nash equilibrium,
in which the payoff for the two strategies of buyers must be the same,
does not exist. This implies that
 Disc must be selected as one of the two buyer's strategies 
in a possible Nash equilibrium.

\textit{Mixture of Buy and Disc: cooperative equilibrium}:
Consider a mixture of Buy and Disc, which we call
the cooperative equilibrium.
Note that
\begin{align}
\MbG= & 1-\epsilon,\\
\MbB= & (1-\epsilon)y_1+\epsilon y_2 = (1-2\epsilon)y_1+\epsilon.
\label{eq:MbB when Buy and Disc coexist}
\end{align}
Because $P_i^{\rm b}(1-\epsilon,1-\epsilon)=P_i^{\rm b}(1-\epsilon,\epsilon)$ must be satisfied
in the Nash equilibrium, the coefficient of $b_{\rm B}$ in 
\EQ\eqref{eq:payoff buyer}
must be equal to 0. This condition combined with
$P^{\rm s}_{\rm C} = P^{\rm s}_{\rm D}$ yields
\begin{equation}
\MbB^* = \frac{(-\mu+rc_1)(1-\epsilon)}{1-\mu-rc_2}
\label{eq:MbB when Buy and Disc coexist 2}
\end{equation}
and
\begin{equation}
x^* = \frac{c_1}{c_1+\mu}.
\label{eq:x* when Buy and Disc coexist}
\end{equation}
Equation~\eqref{eq:x* when Buy and Disc coexist}
replicates \EQ\eqref{eq:x* when Buy and Disc coexist repl}.

The mixture of Buy and Disc is equivalent to
$\epsilon<\MbB^*<1-\epsilon$. The inequality $\MbB^*<1-\epsilon$ is always
satisfied if $r<1$ (note that the denominator of
\EQ\eqref{eq:MbB when Buy and Disc coexist 2} is always positive if $\mu<1/2$).
The inequality $\epsilon<\MbB^*$ is equivalent to \EQ\eqref{eq:Disc necessary cnd 1}.
Given \EQ\eqref{eq:Disc necessary cnd 1}, the equilibrium probability of
Buy (i.e., $y_1^*$) and that of Disc (i.e., $y_2^*=1-y_1^*$)
in the cooperative equilibrium, shown in 
\EQ\eqref{eq:y1* cooperative equilibrium}, are derived by substituting
\EQ\eqref{eq:MbB when Buy and Disc coexist 2} in 
\begin{equation}
\MbB^* = (1-\epsilon)y_1^* + \epsilon y_2^*
= (1-2\epsilon)y_1^* + \epsilon.
\label{eq:MbB* Buy-Disc}
\end{equation}

To show the stability of the cooperative equilibrium,
we first compare
$P_i^{\rm b}(1-\epsilon,1-\epsilon)=P_i^{\rm b}(1-\epsilon,\epsilon)$, i.e., the payoff for
Buy and that for Disc, and $P_i^{\rm b}(\epsilon,\epsilon)$, i.e., the
payoff for NoBuy, in the cooperative equilibrium.
NoBuy gains a smaller payoff than Buy and Disc if the coefficient of
$b_{\rm G}$ in \EQ\eqref{eq:payoff buyer} is
positive in the cooperative equilibrium. The substitution of
\EQS\eqref{eq:MbB when Buy and Disc coexist 2} and
\eqref{eq:x* when Buy and Disc coexist} in \EQ\eqref{eq:payoff buyer}
suggests that this condition is
equivalent to
\begin{equation}
1-\mu>\frac{\mu c_2}{c_1}.
\label{eq:cnd for mu in cooperative equil}
\end{equation}
Equation~\eqref{eq:cnd for mu in cooperative equil} is equivalent to
$\mu<1/2$, which we have assumed. Therefore, 
NoBuy gains a smaller payoff than Buy and Disc. Because 
$P_i^{\rm b}(1-\epsilon,1-\epsilon)=P_i^{\rm b}(1-\epsilon,\epsilon)$
implies
$P_i^{\rm b}(\epsilon,1-\epsilon)=P_i^{\rm b}(\epsilon,\epsilon)$,
AntiDisc also gains a smaller payoff than Buy and Disc in the
cooperative equilibrium.

Consider the invariant subspace of the strategy space where only Buy and Disc
buyers (and C and D sellers) exist. The mixed
strategy specified by $(y_1,x)=(y_1^*,x^*)$ is a Nash equilibrium
when the buyer's strategies are restricted to either Buy and Disc
because $P_i^{\rm b}(1-\epsilon,1-\epsilon)=P_i^{\rm
  b}(1-\epsilon,\epsilon)$ and $P^{\rm s}_{\rm C}=P^{\rm s}_{\rm D}$.
Because $P_i^{\rm b}(\epsilon,1-\epsilon)=P_i^{\rm
  b}(\epsilon,\epsilon)<
P_i^{\rm b}(1-\epsilon,1-\epsilon)=P_i^{\rm
  b}(1-\epsilon,\epsilon)$, $(y_1,x)=(y_1^*,x^*)$ is a Nash
equilibrium when all the four types of buyer's strategies are allowed.

The inequality $P_i^{\rm b}(\epsilon,1-\epsilon)=P_i^{\rm
  b}(\epsilon,\epsilon)<
P_i^{\rm b}(1-\epsilon,1-\epsilon)=P_i^{\rm
  b}(1-\epsilon,\epsilon)$ also assures that, under the replicator dynamics,
the cooperative equilibrium is asymptotically stable against the introduction
of an infinitesimal fraction of AntiDisc or NoBuy.
Therefore, we are left to show the asymptotic stability
of the cooperative equilibrium
within the abovementioned two-dimensional subspace
parametrized by $y_1$ and $x$. For the sake of the linear
stability analysis, we take $\MbB$ and $x$
as the independent variables and linearize
\EQS\eqref{eq:dy/dt} with $i=1$ and
\eqref{eq:dx/dt}
and substitute the one-to-one relationship between $y_1$ and $\MbB$ given by
\EQ\eqref{eq:MbB when Buy and Disc coexist} in 
\EQ\eqref{eq:dy/dt}. The Jacobian in the equilibrium is given by
\begin{equation}
J=\begin{pmatrix}
\frac{\partial}{\partial \MbB}\frac{{\rm d}\MbB}{{\rm d}t}&
\frac{\partial}{\partial x}\frac{{\rm d}\MbB}{{\rm d}t}\\
\frac{\partial}{\partial \MbB}\frac{{\rm d}x}{{\rm d}t}&
\frac{\partial}{\partial x}\frac{{\rm d}x}{{\rm d}t}
\end{pmatrix},
\end{equation}
where all the derivatives are evaluated at
$(\MbB,x)=(\MbB^*,x^*)$, and
\begin{align}
\frac{\partial}{\partial \MbB}\frac{{\rm d}\MbB}{{\rm d}t} =&
(\MbB^*-\epsilon)(1-\epsilon)(1-\epsilon-\MbB^*) \left\{
\frac{-rx^*\mu(1-\mu)}
{[\mu(1-\epsilon)+(1-\mu)\MbB^*]^2} +
\frac{(1-x^*)c_1c_2}{[c_1(1-\epsilon)+c_2\MbB^*]^2}
\right\},\label{eq:J11}\\
\frac{\partial}{\partial x}\frac{{\rm d}\MbB}{{\rm d}t} =&
(\MbB^*-\epsilon)(1-\epsilon)(1-\epsilon-\MbB^*) \left\{
\frac{r\mu}{\mu(1-\epsilon)+(1-\mu)\MbB^*}+
\frac{c_1}{c_1(1-\epsilon)+c_2\MbB^*}
\right\},\label{eq:J12}\\
\frac{\partial}{\partial \MbB}\frac{{\rm d}x}{{\rm d}t} =& 
x^*(1-x^*)(1-\epsilon)\MbB^* \left\{
\frac{-r(1-\mu)}{[\mu(1-\epsilon)+(1-\mu)\MbB^*]^2}
+ \frac{c_2}{[c_1(1-\epsilon)+c_2\MbB^*]^2}
\right\},\label{eq:J21}\\
\frac{\partial}{\partial x}\frac{{\rm d}x}{{\rm d}t} =& 0,
\label{eq:J22}
\end{align}
The necessary and sufficient condition for the cooperative equilibrium
to be stable under the replicator dynamics
is given by ${\rm trace} J<0$ and $\det J>0$.
By substituting
\EQS\eqref{eq:MbB when Buy and Disc coexist 2} and
\eqref{eq:x* when Buy and Disc coexist}
in \EQ\eqref{eq:J11}, we obtain
\begin{equation}
{\rm trace} J = \frac{\mu(\MbB^*-\epsilon)(1-\epsilon-\MbB^*)c_1(1-\mu-rc_2)^2}
{(1-\epsilon)(c_1+\mu)(c_1-\mu)^2}
\left(-\frac{1-\mu}{r}+c_2\right).
\label{eq:trace J}
\end{equation}
Because the right-hand side of \EQ\eqref{eq:J21} is positive, we obtain
\begin{equation}
\det J \propto - \frac{\partial}{\partial \MbB}\frac{{\rm d}x}{{\rm d}t}
= \frac{x^*(1-x^*)\MbB^*(1-\mu-rc_2)^2}{(1-\epsilon)(c_1-\mu)^2}
\left(\frac{1-\mu}{r}-c_2\right).
\label{eq:det J}
\end{equation}
Equations~\eqref{eq:trace J} and \eqref{eq:det J} suggest that the cooperative equilibrium
is stable if and only if
\begin{equation}
\frac{1-\mu}{r} > c_2.
\label{eq:cond coop stable}
\end{equation}
Equation~\eqref{eq:cond coop stable} is satisfied for any $r$ ($0<r<1$) if
$\mu<1/2$. Therefore, the cooperative equilibrium is asymptotically
stable under the replicator dynamics.

\textit{Mixture of Disc and AntiDisc}:
If Disc and AntiDisc are mixed in the equilibrium,
\begin{equation}
P_i^{\rm b}(1-\epsilon,\epsilon)=P_i^{\rm b}(\epsilon,1-\epsilon)
\label{eq:Pb(Disc)=Pb(anti)}
\end{equation}
holds true.
Because $P_i^{\rm b}(b_{\rm G}, b_{\rm B})$ is linear in $b_{\rm G}$ and $b_{\rm B}$,
the coefficient
of $b_{\rm G}$ and that of $b_{\rm B}$ must be the same
for \EQ\eqref{eq:Pb(Disc)=Pb(anti)} to be satisfied. If both coefficients are positive,
the payoff for Buy is larger than that for Disc and AntiDisc
in this equilibrium. If both coefficients are negative,
the payoff for NoBuy is larger than that for Disc and AntiDisc
in the equilibrium. In either case, the mixture of Disc
and AntiDisc cannot be Nash.

\textit{Mixture of Disc and NoBuy}: 
If Disc and NoBuy are mixed in the equilibrium, we obtain
$P_i^{\rm b}(1-\epsilon,\epsilon)=P_i^{\rm b}(\epsilon,\epsilon)$ and
$\MbB=b_{\rm B}=\epsilon$. Therefore, the coefficient of $b_{\rm G}$ in
\EQ\eqref{eq:payoff buyer},
which we denote by $h(y_2)$ as a function of the density of
Disc, is represented as
\begin{equation}
h(y_2)=\frac{rx(1-\mu)\epsilon}{\mu\MbG+(1-\mu)\epsilon} - \frac{(1-x)c_2\epsilon}{c_1\MbG+c_2\epsilon},
\label{eq:h(y2)}
\end{equation}
where
\begin{equation}
\MbG=(1-\epsilon)y_2+\epsilon(1-y_2)
\label{eq:MbG vs y_2 when Disc and NoBuy coexist}
\end{equation}
must be equal to 0 in the equilibrium.
From $h(y_2)=0$ and $P^{\rm s}_{\rm C}=P^{\rm s}_{\rm D}$, we obtain
\begin{equation}
\MbG^* = \frac{1-\mu-rc_2}{-\mu+rc_1}
\end{equation}
and
\begin{equation}
x^* = \frac{c_2}{1-\mu+c_2}.
\end{equation}

If ${\rm d}h(y_2)/{\rm d}y_2>0$ in the equilibrium, 
Disc (NoBuy)
in a mixed strategy in which there are slightly more (less)
probability of Disc
than in the equilibrium 
$(\MbG,x)=(\MbG^*,x^*)$
obtains a larger payoff than NoBuy (Disc). In this case,
$(\MbG^*,x^*)$ is not Nash
because such a slightly modified mixed strategy of the buyer obtains
a larger payoff than $(\MbG^*,x^*)$.
On the basis of the relationship
${\rm d}h/{\rm d}y_2 = (1-2\epsilon){\rm d}h/{\rm d}\MbG$, which is derived from
\EQ\eqref{eq:MbG vs y_2 when Disc and NoBuy coexist},
we rewrite $h(y_2)$ as $h(\MbG)$ and examine
${\rm d}h(\MbG)/{\rm d}\MbG$ in the equilibrium (note that we assumed $\epsilon<1/2$).
Using the fact that the right-hand side of
\EQ\eqref{eq:h(y2)} is equal to 0 when $(\MbG,x)=(\MbG^*,x^*)$, we obtain
\begin{align}
\left.\frac{{\rm d}h(b_{\rm G})}{{\rm d}b_{\rm G}}\right|_{(\MbG,x)=(\MbG^*,x^*)} =&
\frac{-rx^* (1-\mu)\mu\epsilon}{[\mu\MbG^* + (1-\mu)\epsilon]^2}
+ \frac{(1-x^*)c_1c_2\epsilon}
{(c_1\MbG^* + c_2\epsilon)^2}\notag\\
= & \frac{rx^*(1-\mu)(1-2\mu)\epsilon^2\theta}
{[\mu\MbG^* + (1-\mu)\epsilon]^2 (c_1\MbG^* + c_2\epsilon)}\notag\\
>& 0
\end{align}
if $\mu<1/2$. Therefore, the mixture of Disc and NoBuy is
not Nash.

\subsubsection*{Mixture of three or four buyer's strategies}

If three or four buyer's strategies are mixed in an equilibrium,
their payoffs must be identical. Therefore, the coefficient of $b_{\rm
G}$ and that of $b_{\rm B}$ in \EQ\eqref{eq:payoff buyer} 
must be equal to 0, which requires
$\theta (2\mu - 1)=0$. Because $\mu<1/2$,
this relationship is not satisfied when the image scorer
exists (i.e., $\theta> 0$).

When $\theta=0$, we obtain
$P^{\rm s}_{\rm C}<P^{\rm s}_{\rm D}$ by substituting $c_1=\mu$ and
$c_2=1-\mu$ (derived from \EQS\eqref{eq:c1} and \eqref{eq:c2})
in \EQS\eqref{eq:payoff C seller} and \eqref{eq:payoff D
seller}. Therefore, $x^*=0$ must hold in the equilibrium.
In this situation, \EQ\eqref{eq:payoff buyer} is reduced to
\begin{equation}
P_i^{\rm b}(b_{\rm G}, b_{\rm B}) = -
\frac{(1-\mu)b_{\rm G}\MbB+\mu\MbG b_{\rm B}}
{\mu\MbG+(1-\mu)\MbB}.
\label{eq:payoff buyer 3 or 4 strategies}
\end{equation}
Equation~\eqref{eq:payoff buyer 3 or 4 strategies} indicates that
NoBuy's payoff is larger than Disc's and AntiDisc's payoffs, which are larger
than Buy's payoff.
Consequently, three or four buyers' strategies cannot be mixed
in an equilibrium.
Here is the end of the proof.

\subsection*{Indifferent scoring}\label{sub:indifferent}

In a case with only indifferent scorers (i.e., $\theta=0$), 
\EQ\eqref{eq:Disc necessary cnd 1 replicated} is never satisfied
(and \EQ\eqref{eq:Disc singular main text} is not satisfied, either).
Therefore, the uncooperative equilibrium is the only
Nash equilibrium.
This outcome is expected because, under indifferent scoring,
the players perform the usual trust game
\cite{Dasgupta1988book,Kreps1990chap,Berg1995geb,Fehr2003Nature}.

\subsection*{Image scoring}\label{sub:image}

When there are only image scorers (i.e., $\theta=1$),
the cooperative equilibrium is
realized in a wide parameter region because \EQ\eqref{eq:Disc necessary cnd 1 replicated} with $\theta=1$ and $\mu\to 0$ is reduced to
$\epsilon/(1-\epsilon)<r<1$; $\epsilon<1/2$ is the probability that
the buyer misimplements the action.  In the limit $\mu\to 0$, the
equilibrium probability of Buy
(i.e., $y_1^*$) is plotted as a function of $\epsilon$ and $r$ in
\FIG\ref{fig:y1*}(a).  In the parameter region in which $y_1^*=0$
(black region in \FIG\ref{fig:y1*}(a)), the cooperative equilibrium
does not exist.
In the cooperative equilibrium, the fraction of Buy is
large for a large $r$ or a small $\epsilon$.  In particular,
$\lim_{\mu\to 0, r\to 1}y_1^*=1$ irrespective of the value of
$\epsilon$. In the limit $r\to 1$, the trust game is a weak social dilemma such that
the D seller's payoff is only infinitesimally larger than
the C seller's payoff. 
The advantage of the D seller
is offset by the B reputation that the
D seller receives from just a small fraction of Disc buyers (i.e., $y_2\ll 1$).
Even if a majority of buyers is nondiscriminative Buy, cooperation between
buyers and sellers can be sustained by the reputation-regarding
behavior of a small fraction of Disc.

For a general value of $\theta$,
suppose that \EQ\eqref{eq:Disc necessary cnd 1 replicated}
is satisfied such that the cooperative equilibrium is Nash.
For $\epsilon=0.1$ and $\mu\to 0$,
$y_1^*$ in the cooperative equilibrium
is shown for various values of $\theta$ and $r$ in
\FIG\ref{fig:y1*}(b). $y_1^*$ increases as
$\theta$ and $r$ increase.

For some values of $\mu$ and $\epsilon$, the critical line
obtained from \EQ\eqref{eq:Disc necessary cnd 1 replicated} is plotted
in \FIG\ref{fig:r-crit-theta}. The cooperative equilibrium exists and
is stable above the critical line.
Figure~\ref{fig:r-crit-theta} indicates that cooperation based on
the reputation mechanism occurs in a large parameter region, particularly
for large values of $\theta$ and $r$. The critical line is rather
insensitive to $\mu$ and relatively sensitive to
$\epsilon$. Nevertheless, cooperation is possible even for
a large probability of the implementation error $\epsilon=0.2$.

\subsection*{Attractive basin of the cooperative and uncooperative equilibria}

The cooperative and uncooperative equilibria can coexist.
A fuller understanding of the model requires a global analysis
of the replicator dynamics to determine which of the two equilibria is more likely to be attained.
In addition, periodic or chaotic attractors may exist when a population evolves.
To exclude this possibility and examine the attractive basin
of the two equilibria, we numerically run the standard two-population
replicator dynamics \cite{Weibull1995book,Hofbauer1998book,Gintis2009book_evolving}
from various initial conditions
to identify the limit set of the dynamics.
For fixed parameter values, we assume that
the initial condition is distributed according to the uniform density on the
state space, i.e., $\{ (y_1, y_2, y_3, y_4, x) : y_i\ge 0$ $(1\le
i\le 4), \sum_{i=1}^4 y_i=1, 0\le x\le 1\}$. 
It should be noted that \EQ\eqref{eq:yiIN=yiIM} is preserved
under the replicator dynamics
because the buyer's payoff depends on the buyer's strategy 
but is independent of whether the buyer is an indifferent or image scorer.
We have implicitly assumed in \EQS\eqref{eq:dy/dt} 
and \eqref{eq:dx/dt}
that sellers and buyers have identical
adaptation rates.  However, the following results are qualitatively
the same even if the two adaptation rates are different.

The replicator dynamics are four-dimensional, with
three degrees of freedom derived from
the buyer's population and
one degree of freedom derived from the seller's population.
Note
that the selection occurs separately among buyers and sellers.

For an expository purpose, we start with the system without
AntiDisc. There are three strategies for buyers and two strategies for
sellers. The replicator dynamics are three-dimensional, and the state
space is a triangular prism defined by $\{ (y_1,y_2,y_4,x) :
y_1, y_2,
y_4\ge 0, y_3=0, y_1+y_2+y_4=1, 0\le x\le 1 \}$. For $\mu=0.02$, $\epsilon=0.1$,
$r=0.15$, and $\theta=1$, initial conditions located
below the boundary shown in \FIG\ref{fig:triangular prism} are attracted to the uncooperative equilibrium.
We confirmed that all the complementary regions in the interior of the triangular
prism are attracted to the cooperative equilibrium.
Our numerical simulations strongly suggest that there is no
other limit set.

To better quantify the possibility of cooperation,
we measure the volume of the attractive basin
of the cooperative equilibrium. The relative volume of the
attractive basin in the triangular prism
is shown in \FIG\ref{fig:volume r-theta}(a)
for $\mu=0.02$, $\epsilon=0.1$, and
various values of $r$ and $\theta$.
The critical line for the existence and stability of the
cooperative equilibrium,
implied by
\EQ\eqref{eq:Disc necessary cnd 1 replicated}, is 
shown by the solid line.
We find that the cooperative equilibrium is attractive for a substantial
variety of initial conditions as $\theta$ and $r$ increase.

In the presence of four strategies for buyers, 
the relative volume of the attractive basin of the
cooperative equilibrium in the state space is shown in
\FIG\ref{fig:volume r-theta}(b).  The results are qualitatively the
same as those shown in \FIG\ref{fig:volume r-theta}(a). We confirmed
that the rest of the state space belongs to the attractive basin of
the uncooperative equilibrium. 
The introduction of AntiDisc does not
inhibit the evolution of cooperation.

\section*{Discussion}\label{sec:discussion}

For the trust game, we have shown that cooperation between 
unacquainted buyers and sellers
can be established under the image scoring norm (i.e., reputation mechanism).
In the cooperative equilibrium, the population of buyers (i.e., investors)
is a mixture of
Buy (unconditional buyer) and Disc (discriminator that decides whether to buy
depending on the seller's reputation). The majority of
sellers (i.e., trustees) reciprocates the buyer's trust although the sellers are not expected
to meet the same buyers again. It should be noted that not every buyer discriminates between
good and bad sellers in the cooperative equilibrium. This feature 
is shared by some previous models of indirect reciprocity
\cite{Panchanathan2003jtb,Brandt2006jtb,OhtsukiIwasa2007jtb}.
The probability of discriminators can be small depending on
the parameter values.
In addition, the buyer's reputation may be of little practical use for
maintaining cooperative transactions.
This claim is consistent with the previous theoretical result
\cite{Klein1981JPE} and the empirical finding that
reputation for a seller
has a greater impact than that for a buyer
\cite{Malaga2001ECR,Resnick2002AAM,Brown2006California}.

Our model and results are distinct from previous ones obtained from the
models using the symmetrized donation game, which we call
indirect reciprocity games for now
\cite{Nowak1998nature,Nowak1998jtb,Leimar2001RoyalB,Panchanathan2003jtb,OhtsukiIwasa2004jtb,Brandt2004jtb,Nowak2005nature,Sigmund2010book}.
First, when cooperation prevails, a player with a good reputation is
helped by unacquainted players in the indirect reciprocity games.  In
our model, a seller with a good reputation wins the trust, not
explicit help, of unacquainted buyers.  By reciprocating the buyer's trust,
the seller obtains a relatively large momentary payoff and a
good reputation. Then, a good reputation elicits trust and long-term
cooperation from buyers.

A second difference is in the consequence of the image scoring.
In the indirect reciprocity games,
defection against a bad player is regarded to be
bad under the image scoring. Therefore, the image scoring does not yield
cooperation
\cite{Leimar2001RoyalB,Panchanathan2003jtb,OhtsukiIwasa2004jtb,Brandt2004jtb}.
In contrast, in our model, a discriminative buyer (i.e., Disc) that defects against
(i.e., does not buy from) a bad seller does not receive a bad
reputation. This is because buyers do not own reputation scores by definition;
the asymmetry of roles in the trust game allows the image scoring to support cooperation.
Although a previous paper discussed this issue before
\cite{Mcnamara2009RoyalB}, we analytically derived the conditions under which
cooperation based on the image scoring occurs.
It should be noted that the image scoring norm is
simpler than the social norms required for cooperation in the indirect
reciprocity games
\cite{Leimar2001RoyalB,Panchanathan2003jtb,OhtsukiIwasa2004jtb,OhtsukiIwasa2006jtb,OhtsukiIwasa2007jtb,Brandt2004jtb}.

In other models, the 
mere reputation mechanism enables cooperation among unacquainted players in the trust
game \cite{Sigmund2001PNAS,Sigmund2010book} and other games
\cite{Raub1990AJS,Kandori1992RES,Nowak2000Science,Sigmund2010book}.
However, the component social dilemma game in
these studies is essentially symmetric. Our model is inherently asymmetric such that
a player is either a permanent buyer or permanent seller.
As compared to a recent paper in which impacts of
asymmetric roles in the trust game are numerically studied \cite{Mcnamara2009RoyalB},
we analytically established the conditions under which
reputation-based cooperation occurs in the asymmetric trust game.
  In online
marketplaces \cite{Resnick2002AAM}, the market for lemons
\cite{Akerlof1970QJE}, and presumably many other transaction scenes, some people
may participate entirely or mostly as buyers and others as sellers.

The constructive role of reputations for the trustee (i.e., seller) 
in asymmetric interactions was
formulated in a classic paper many years ago \cite{Klein1981JPE}.
Our contribution to the understanding of this established mechanism
is that we have clarified competition among different strategies
using evolutionary game theory.

We have investigated a scenario in which indifferent scorers, which
do not essentially score sellers, and
image scorers coexist in a buyer's population.  Remarkably,
the fraction of image scorers needed for cooperation
is not large; \EQ\eqref{eq:Disc necessary cnd 1 replicated} indicates that 
the threshold fraction of image scorers is equal to 0.60 for $r=0.2$ and 0.038 for $r=0.8$
when $\mu=0.02$ and $\epsilon=0.1$.
An alternative assumption for the behavior of indifferent
scorers is that they do not alter sellers' scores rather than always give a good reputation to sellers. Analysis of this case is warranted for future work.

An important limitation of the present study is that
the fraction of indifferent scorers and that of image scorers are invariant
over time. In other words,
indifferent and image scorers are assumed to receive the same
payoff if the strategy (Buy, Disc, AntiDisc, or NoBuy) is the same.
In fact, the image scoring may be more costly than the indifferent
scoring because the image scorer has to know and report
whether sellers cooperate or defect. 
Therefore, an image scorer may be
tempted to turn into an indifferent scorer if
the incentive to rate sellers
is absent \cite{Resnick2000ACM,Resnick2002AAM,Malaga2001ECR}. 
A similar cost is briefly mentioned in previous literature in 
the context of indirect reciprocity games \cite{Milinski2001royalb,Sigmund2010book}. In our model, cooperation disappears if
there are too many indifferent scorers. 
This result parallels with that for the indirect reciprocity games in which
cooperation is not realized if there are too few observers in the one-shot game
\cite{Nowak1998nature,Brandt2004jtb}.
In practice, 
rewarding image scorers and
shutting down indifferent scorers' access to reputation information, for example,
are means to circumvent the scoring cost
\cite{Malaga2001ECR}.
We remark that the competition of social norms was analyzed
in different models of indirect reciprocity
 \cite{Chalub2006jtb,Pacheco2006PlosComputBiol,Uchida2010jtb}.

In the context of asymmetric interaction between cleaner and client
fishes, indirect reciprocity was investigated in an inherently asymmetric
variant of the trust
game with a binary internal state for the trustee
\cite{Johnstone2007royalb}.  Our model and results are distinct from
theirs. In their model, a trustee reciprocates and attracts the
investor when the trustee is in one particular state and exploits the
investor by switching to the other state. The trustee does not
steadily maintain a good reputation and uses the temporarily good
reputation to exploit the investor. The authors acknowledge that their
mechanism is different from the
conventional concept of indirect reciprocity.  In our model, the
trustee (i.e., seller) steadily maintains a good reputation to invoke
help from the investor (i.e., buyer). Our results suggest that, under
appropriate conditions, the conventional indirect reciprocity may be
established between cleaner and client fishes without resorting to the
concept of the binary state.  Finally, beyond the relevance to
online transactions, our results
provide a firm solution to the moral hazard problem
that is represented by the trust game.
Examples include offline
markets \cite{Klein1981JPE} 
such as the market for lemons \cite{Akerlof1970QJE} and
labor markets \cite{Kreps1990chap,McCabe2003JEBO}. Implementing a reputation mechanism
only for the seller (i.e., investee) induces trust and cooperation 
between unacquainted individuals in the trust game. Good cooperative 
sellers and trustful
buyers coevolve.

\section*{Acknowledgments}

We thank Michihiro Kandori for discussion and Hisashi Ohtsuki for providing
valuable comments on the manuscript.

% PLoS ONE
% N.M. acknowledges the support through Grants-in-Aid for Scientific Research (No.\ 20760258) from MEXT, Japan. M.N. acknowledges the support and Grants-in-Aid for Scientific Research from JSPS, Japan.

\newpage
\clearpage

\newcounter{figcount}
\renewcommand{\thefigcount}{\arabic{figcount}}
\newcommand{\figcount}{\refstepcounter{figcount}%
           Figure~\thefigcount:~}

\section*{Figure captions}

\noindent
\figcount\label{fig:schem trust game}
Schematic of the trust game.

\bigskip

\noindent
\figcount\label{fig:y1*}
Probability of Buy $y_1^*$ in the cooperative equilibrium in the limit
$\mu\to 0$. We set (a) $\theta=1$ and (b) $\epsilon=0.1$.

\bigskip

\noindent
\figcount\label{fig:r-crit-theta}
The threshold value of $r$ above which the cooperative
equilibrium exists and is stable.

\bigskip

\noindent
\figcount\label{fig:triangular prism}
Boundary between the attractive basins of the cooperative and 
uncooperative equilibria. The points above the boundary are attracted to the
cooperative equilibrium. We set
$\mu=0.02$, $\epsilon=0.1$, $r=0.15$, and $\theta=1$.

\bigskip

\noindent
\figcount\label{fig:volume r-theta}
Relative volume of the attractive basin of the cooperative
  equilibrium. We set $\mu=0.02$ and $\epsilon=0.1$. Initially,
(a) three and (b) four buyer's strategies are distributed
  according to the uniform density.

\newpage

%\begin{figure}
\begin{center}
\includegraphics[width=8cm]{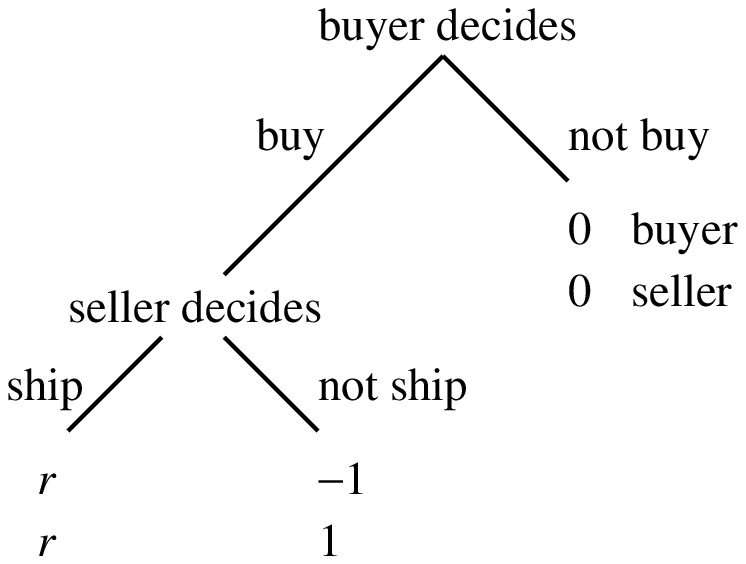}\\
%\caption{Schematic of the trust game.}
%\label{fig:schem trust game}
\bigskip
\textbf{Figure 1}
\end{center}
%\end{figure}

\clearpage

%\begin{figure}
\begin{center}
\includegraphics[width=15cm]{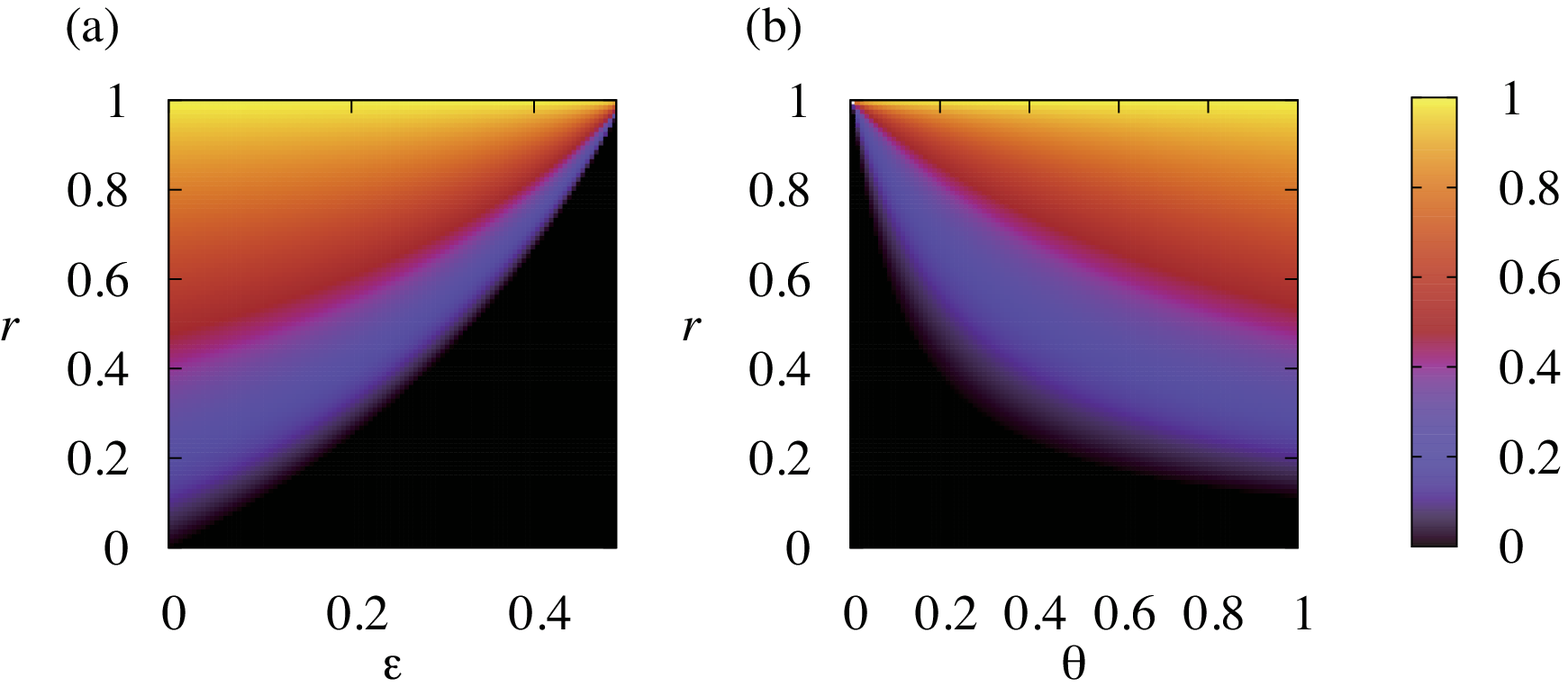}\\
%\includegraphics[width=8cm]{y1-equilibrium-theta1}
%\includegraphics[width=8cm]{y1-equilibrium-eps01}\\
%\caption{Probability of Buy $y_1^*$ in the cooperative equilibrium in the limit
%$\mu\to 0$. We set (a) $\theta=1$ and (b) $\epsilon=0.1$.}
%\label{fig:y1*}
\bigskip
\textbf{Figure 2}
\end{center}
%\end{figure}

\clearpage

%\begin{figure}
\begin{center}
\includegraphics[width=8cm]{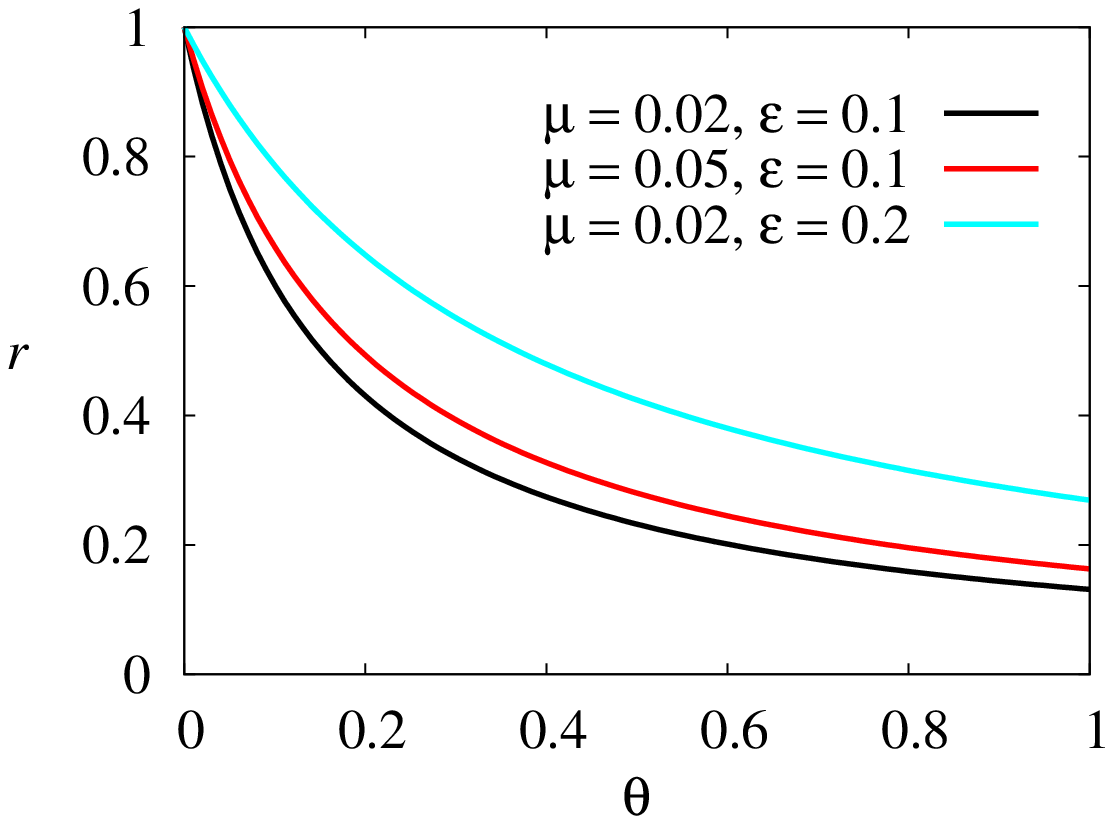}\\
%\caption{The threshold value of $r$ above which the cooperative
%equilibrium exists and is stable.}
%\label{fig:r-crit-theta}
\bigskip
\textbf{Figure 3}
\end{center}
%\end{figure}

\clearpage

%\begin{figure}
\begin{center}
\includegraphics[width=8cm]{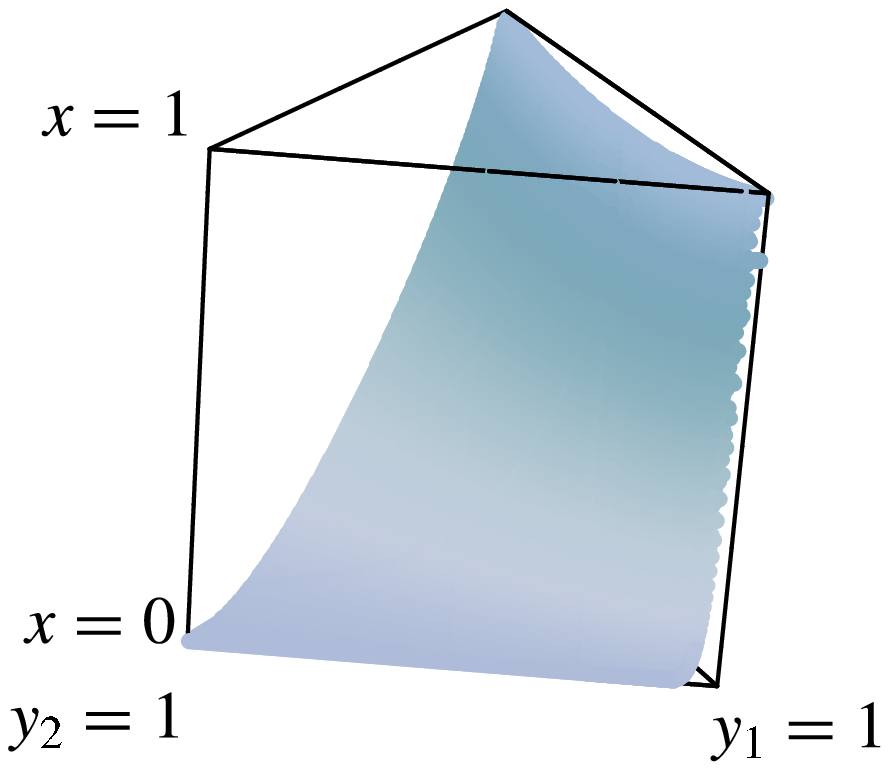}\\
%\caption{Boundary between the attractive basins of the cooperative and 
%uncooperative equilibria. The points above the boundary are attracted to the
%cooperative equilibrium. We set
%$\mu=0.02$, $\epsilon=0.1$, $r=0.15$, and $\theta=1$.}
%\label{fig:triangular prism}
\bigskip
\textbf{Figure 4}
\end{center}
%\end{figure}

\clearpage

%\begin{figure}
\begin{center}
\includegraphics[width=15cm]{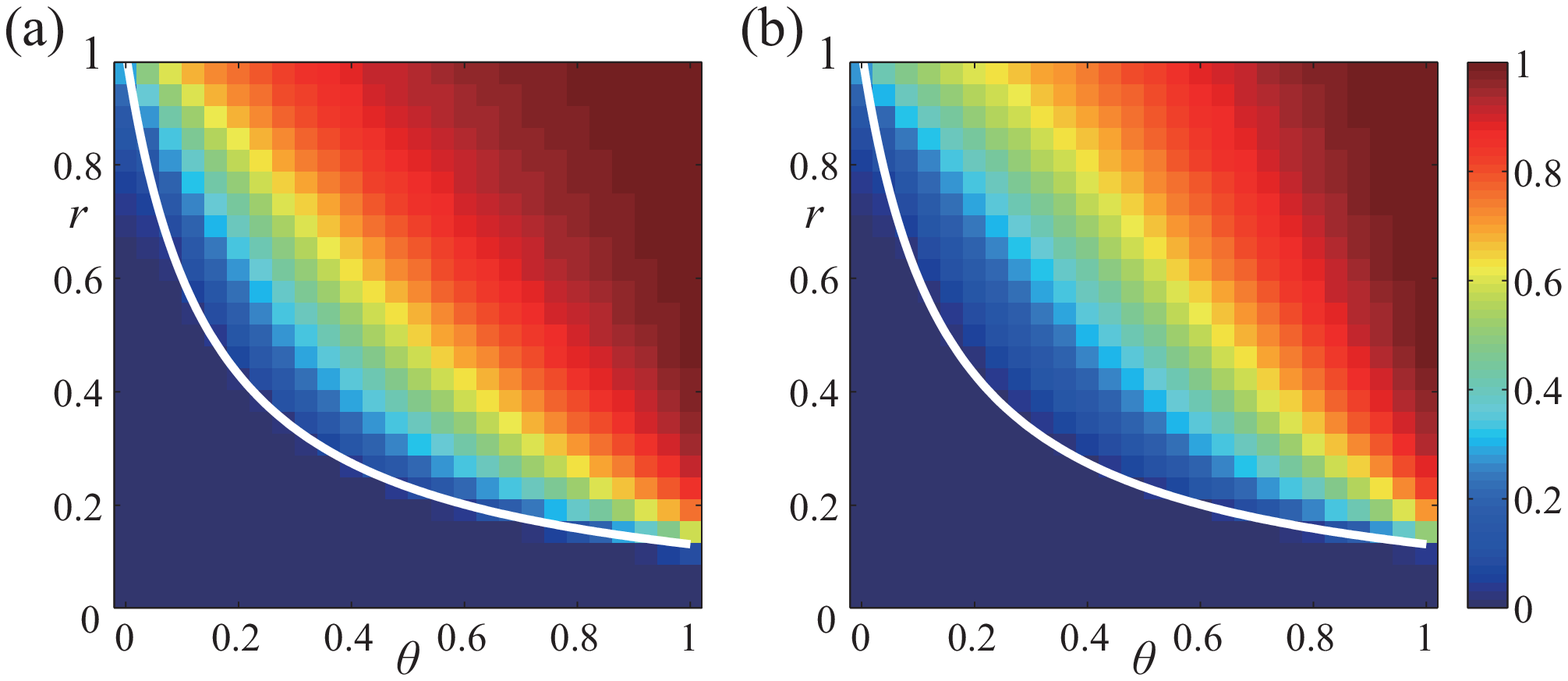}\\
%\caption{Relative volume of the attractive basin of the cooperative
%  equilibrium. We set $\mu=0.02$ and $\epsilon=0.1$. Initially,
%(a) three and (b) four buyer's strategies are distributed
%  according to the uniform density.}
%\label{fig:volume r-theta}
\bigskip
\textbf{Figure 5}
\end{center}
%\end{figure}

%\end{document}


\begin{thebibliography}{10}
\providecommand{\url}[1]{\texttt{#1}}
\providecommand{\urlprefix}{URL }
\expandafter\ifx\csname urlstyle\endcsname\relax
  \providecommand{\doi}[1]{doi:\discretionary{}{}{}#1}\else
  \providecommand{\doi}{doi:\discretionary{}{}{}\begingroup
  \urlstyle{rm}\Url}\fi
\providecommand{\bibAnnoteFile}[1]{%
  \IfFileExists{#1}{\begin{quotation}\noindent\textsc{Key:} #1\\
  \textsc{Annotation:}\ \input{#1}\end{quotation}}{}}
\providecommand{\bibAnnote}[2]{%
  \begin{quotation}\noindent\textsc{Key:} #1\\
  \textsc{Annotation:}\ #2\end{quotation}}
\providecommand{\eprint}[2][]{\url{#2}}

\bibitem{Akerlof1970QJE}
Akerlof GA (1970) The market for ``lemons'': Quality uncertainty and the market
  mechanism.
\newblock Quarterly J Econom 84: 488--500.
\input{Akerlof1970QJE}

\bibitem{Resnick2000ACM}
Resnick P, Zeckhauser R, Friedman E, Kuwabara K (2000) Reputation systems.
\newblock Communications of the ACM 43: 45--48.
\input{Resnick2000ACM}

\bibitem{Malaga2001ECR}
Malaga RA (2001) Web-based reputation management systems: Problems and
  suggested solutions.
\newblock Electronic Commerce Res 1: 403--417.
\input{Malaga2001ECR}

\bibitem{Dellarocas2003ManSci}
Dellarocas C (2003) The digitization of word of mouth: Promise and challenges
  of online feedback mechanisms.
\newblock Manage Sci 49: 1407--1424.
\input{Dellarocas2003ManSci}

\bibitem{Brown2006California}
Brown J, Morgan J (2006) Reputation in online auctions: the market for trust.
\newblock California Man Rev 49: 61--81.
\input{Brown2006California}

\bibitem{Klein1981JPE}
Klein B, Leffler KB (1981) The role of market forces in assuring contractual
  performance.
\newblock J Polit Economy 89: 615--641.
\input{Klein1981JPE}

\bibitem{Bolton2004ManSci}
Bolton GE, Katok E, Ockenfels A (2004) How effective are electronic reputation
  mechanisms? an experimental investigation.
\newblock Manage Sci 50: 1587--1602.
\input{Bolton2004ManSci}

\bibitem{Nowak2005nature}
Nowak MA, Sigmund K (2005) Evolution of indirect reciprocity.
\newblock Nature 437: 1291--1298.
\input{Nowak2005nature}

\bibitem{Sigmund2010book}
Sigmund K (2010) The Calculus of Selfishness.
\newblock Princeton, NJ: Princeton University Press.
\input{Sigmund2010book}

\bibitem{Resnick2002AAM}
Resnick P, Zeckhauser R (2002) Trust among strangers in internet transactions:
  Empirical analysis of ebay's reputation system.
\newblock Adv Appl Microeconomics 11: 127--157.
\input{Resnick2002AAM}

\bibitem{Nowak1998nature}
Nowak MA, Sigmund K (1998) Evolution of indirect reciprocity by image scoring.
\newblock Nature 393: 573--577.
\input{Nowak1998nature}

\bibitem{Nowak1998jtb}
Nowak MA, Sigmund K (1998) The dynamics of indirect reciprocity.
\newblock J Theor Biol 194: 561--574.
\input{Nowak1998jtb}

\bibitem{Leimar2001RoyalB}
Leimar O, Hammerstein P (2001) Evolution of cooperation through indirect
  reciprocity.
\newblock Proc R Soc B 268: 745--753.
\input{Leimar2001RoyalB}

\bibitem{Panchanathan2003jtb}
Panchanathan K, Boyd R (2003) A tale of two defectors: the importance of
  standing for evolution of indirect reciprocity.
\newblock J Theor Biol 224: 115--126.
\input{Panchanathan2003jtb}

\bibitem{OhtsukiIwasa2004jtb}
Ohtsuki H, Iwasa Y (2004) How should we define goodness?--reputation dynamics
  in indirect reciprocity.
\newblock J Theor Biol 231: 107--120.
\input{OhtsukiIwasa2004jtb}

\bibitem{Brandt2004jtb}
Brandt H, Sigmund K (2004) The logic of reprobation: assessment and action
  rules for indirect reciprocation.
\newblock J Theor Biol 231: 475--486.
\input{Brandt2004jtb}

\bibitem{Brandt2006jtb}
Brandt H, Sigmund K (2006) The good, the bad and the discriminator --- errors
  in direct and indirect reciprocity.
\newblock J Theor Biol 239: 183--194.
\input{Brandt2006jtb}

\bibitem{OhtsukiIwasa2006jtb}
Ohtsuki H, Iwasa Y (2006) The leading eight: social norms that can maintain
  cooperation by indirect reciprocity.
\newblock J Theor Biol 239: 435--444.
\input{OhtsukiIwasa2006jtb}

\bibitem{OhtsukiIwasa2007jtb}
Ohtsuki H, Iwasa Y (2007) Global analyses of evolutionary dynamics and
  exhaustive search for social norms that maintain cooperation by reputation.
\newblock J Theor Biol 244: 518--531.
\input{OhtsukiIwasa2007jtb}

\bibitem{Chalub2006jtb}
Chalub FACC, Santos FC, Pacheco JM (2006) The evolution of norms.
\newblock J Theor Biol 241: 233--240.
\input{Chalub2006jtb}

\bibitem{Pacheco2006PlosComputBiol}
Pacheco JM, Santos FC, Chalub FACC (2006) Stern-judging: A simple, successful
  norm which promotes cooperation under indirect reciprocity.
\newblock PLoS Comput Biol 2: 1634--1638.
\input{Pacheco2006PlosComputBiol}

\bibitem{Uchida2010jtb}
Uchida S, Sigmund K (2010) The competition of assessment rules for indirect
  reciprocity.
\newblock J Theor Biol 263: 13--19.
\input{Uchida2010jtb}

\bibitem{Johnstone2007royalb}
Johnstone RA, Bshary R (2007) Indirect reciprocity in asymmetric interactions:
  When apparent altruism facilitates profitable exploitation.
\newblock Proc R Soc B 274: 3175--3181.
\input{Johnstone2007royalb}

\bibitem{Nowak2000Science}
Nowak MA, Page KM, Sigmund K (2000) Fairness versus reason in the ultimatum
  game.
\newblock Science 289: 1773--1775.
\input{Nowak2000Science}

\bibitem{Sigmund2001PNAS}
Sigmund K, Hauert C, Nowak MA (2001) Reward and punishment.
\newblock Proc Natl Acad Sci USA 98: 10757--10762.
\input{Sigmund2001PNAS}

\bibitem{Raub1990AJS}
Raub W, Weesie J (1990) Reputation and efficiency in social interactions: an
  example of network effects.
\newblock Am J Sociol 96: 626--654.
\input{Raub1990AJS}

\bibitem{Kandori1992RES}
Kandori M (1992) Social norms and community enforcement.
\newblock Rev of Econom Studies 59: 63--80.
\input{Kandori1992RES}

\bibitem{Dasgupta1988book}
Dasgupta P (1988) Trust as a commodity.
\newblock In: Trust: making and breaking cooperative relations, D Gambetta, ed
  : 49--72.
\input{Dasgupta1988book}

\bibitem{Kreps1990chap}
Kreps DM (1990) Cooporate culture and economic theory.
\newblock In: Perspectives on Positive Political Economy (J Alt and K Shepsle,
  Eds) : 90--143.
\input{Kreps1990chap}

\bibitem{Berg1995geb}
Berg J, Dickhaut J, McCabe K (1995) Trust, reciprocity, and social history.
\newblock Games Econom Behav 10: 122--142.
\input{Berg1995geb}

\bibitem{Fehr2003Nature}
Fehr E, Fischbacher U (2003) The nature of human altruism.
\newblock Nature 425: 785--791.
\input{Fehr2003Nature}

\bibitem{McCabe2000pnas}
McCabe KA, Smith VL, LePore M (2000) Intentionality detection and
  ``mindreading'': Why does game form matter?
\newblock Proc Natl Acad Sci USA 97: 4404--4409.
\input{McCabe2000pnas}

\bibitem{McCabe2003JEBO}
McCabe KA, Rigdon ML, Smith VL (2003) Positive reciprocity and intentions in
  trust games.
\newblock J Econom Behav {\&} Organization 52: 267--275.
\input{McCabe2003JEBO}

\bibitem{Keser2003IBM}
Keser C (2003) Experimental games for the design of reputation management
  systems.
\newblock IBM Syst J 42: 498--506.
\input{Keser2003IBM}

\bibitem{Basu2009pnas}
Basu S, Dickhaut J, Hecht G, Towry K, Waymire G (2009) Recordkeeping alters
  economic history by promoting reciprocity.
\newblock Proc Natl Acad Sci USA 106: 1009--1014.
\input{Basu2009pnas}

\bibitem{Bracht2009jpe}
Bracht J, Feltovich N (2009) Whatever you say, your reputation precedes you:
  Observation and cheap talk in the trust game.
\newblock J Public Econom 93: 1036-1044.
\input{Bracht2009jpe}

\bibitem{Greif1993AER}
Greif A (1993) {Contract enforceability and economic institutions in early
  trade: the Maghribi traders' coalition}.
\newblock Amer Econ Rev 83: 525--548.
\input{Greif1993AER}

\bibitem{Greif2006book}
Greif A (2006) Institutions and the path to the modern economy.
\newblock Cambridge: Cambridge University Press.
\input{Greif2006book}

\bibitem{Diekmann2005ESSA}
Diekmann A, Przepiorka W (2005) The evolution of trust and reputation: results
  from simulation experiments.
\newblock Third ESSA Conference : 1--7.
\input{Diekmann2005ESSA}

\bibitem{Resnick2006ExpEcon}
Resnick P, Zeckhauser R, Swanson J, Lockwood K (2006) The value of reputation
  on ebay: A controlled experiment.
\newblock Exp Econ 9: 79--101.
\input{Resnick2006ExpEcon}

\bibitem{Bravo2008Rational}
Bravo G, Tamburino L (2008) The evolution of trust in non-simultaneous exchange
  situations.
\newblock Rationality and Society 20: 85--113.
\input{Bravo2008Rational}

\bibitem{Samuelson1992JET}
Samuelson L, Zhang J (1992) Evolutionary stability in asymmetric games.
\newblock J Econom Theory 57: 363--391.
\input{Samuelson1992JET}

\bibitem{Weibull1995book}
Weibull JW (1995) Evolutionary Game Theory.
\newblock Cambridge, MA: MIT Press.
\input{Weibull1995book}

\bibitem{Hofbauer1998book}
Hofbauer J, Sigmund K (1998) Evolutionary Games and Population Dynamics.
\newblock Cambridge, UK: Cambridge University Press.
\input{Hofbauer1998book}

\bibitem{Gintis2009book_evolving}
Gintis H (2009) Game Theory Evolving, Second Edition.
\newblock Princeton, NJ: Princeton University Press.
\input{Gintis2009book_evolving}

\bibitem{Mcnamara2009RoyalB}
McNamara JM, Stephens PA, Dall SRX, Houston AI (2009) {Evolution of trust and
  trustworthiness: social awareness favours personality differences}.
\newblock Proc R Soc B 276: 605--613.
\input{Mcnamara2009RoyalB}

\bibitem{Milinski2001royalb}
Milinski M, Semmann D, Bakker TCM, Krambeck HJ (2001) Cooperation through
  indirect reciprocity: Image scoring or standing strategy?
\newblock Proc R Soc B 268: 2495--2501.
\input{Milinski2001royalb}

\end{thebibliography}
\end{document}